\documentclass[a4paper, 11pt]{article}

\pdfoutput=1

\def\b{\begin{eqnarray}}
\def\e{\end{eqnarray}}
\def\n{\noindent}

\begin{document}

\begin{center}
{\huge \textbf{Evaporation Screening, Scattering,  \vskip.23cm  and Quantum Tunneling Near the \vskip.2cm Horizons of Schwarzschild and
\vskip.55cm Reissner--Nordstr\"om Black Holes}}

\vspace {10mm}
\noindent
{\large \bf Emil M. Prodanov} \vskip.5cm
{\it School of Mathematical Sciences, Technological University Dublin, \vskip.1cm Kevin Street Campus, Dublin 8,
Ireland,} \vskip.1cm
{\it E-Mail: emil.prodanov@dit.ie} \\
\vskip1cm
\end{center}

\vskip4cm
\begin{abstract}
\n
For the radial motion of massive particles with large angular momenta in Schwarzschild geometry and that of massive charged particles with large
angular momenta or energy in a particular range in Reissner--Nordstr\"om geometry, there exist classically forbidden regions on the outside of
the respective event horizons which scatter certain infalling geodesics or screen some of the black holes' evaporation by reflecting the emitted
particles back into the black holes. Quantum  tunneling across this forbidden regions is studied.

\end{abstract}

\newpage

\section{Introduction}

Quantum tunneling in general-relativistic set-up has been an area of interest recently. Parikh and Wilczek \cite{pw} introduced the idea to view
Hawking radiation \cite{hawking} as a tunneling process, based on particles in a dynamical geometry --- such across-horizon phenomena arise as
result of a dynamical fluctuation of the black hole's mass $M$, i.e. a shell of energy $\omega$ moving on the geodesics and replacing $M$ by
$M - \omega$. An outgoing particle starting just inside the initial position of the horizon at $r = 2M - \delta$ (where $\delta$ is a small
positive number) experiences a classically forbidden region and materializes just outside of the final position of the horizon at
$r = 2(M - \omega) + \delta$. \\
Tunneling of massless and massive particles through a smeared quantum horizon has also been examined in the case of black holes in
non-commutative
geometry by considering the effect of smearing of the particle's mass as a Gaussian profile in a flat space-time \cite{nm0} or by extending the
Parikh--Wilczek scheme to higher dimensional non-commutative geometry \cite{nm}.  \\
Quantum tunneling in a naked singularity setting in Reissner--Nordstr\"om geometry in the context of daemon decay has also been studied
\cite{ep}. \\
In Schwarzschild geometry and in Reissner--Nordstr\"om geometry (and, more generally, in Kerr--Newman geometry), the effective one-dimensional
radial motion of massive (charged or not) particles (with non-zero angular momentum) has turning points which bound classically forbidden
regions (in which the kinetic energy is negative). Classically, motion can never occur in such regions. On the other hand, there can never be a
horizon within a classically forbidden region and thus, a horizon cannot get "hidden" inside a potential barrier and in result allow a particle
to tunnel quantum-mechanically across it. The forbidden regions are either on the inside of the inner horizon or on the outside of the outer
one and the loci of the turning points depend on the first integrals of the particle in the corresponding geometry, together with the charge
and mass of the probe. \\
For a marginally bound particle (infalling from state of rest at infinity) or for a higher-energy unbound particle in Schwarzschild geometry,
the
forbidden region is on the outside of the event horizon and it grows towards the event horizon and extends unboundedly with the increase of the
probe's angular momentum. The geodesics of such infalling particles would be scattered and the black hole would be completely "shielded" behind
the forbidden region. In a similar way, the Hawking radiation from a Schwarzschild black hole (and any asymptotically flat black hole) is
dominated by low angular momentum modes --- the higher the angular momentum of the particle, the closer to the event horizon the screen would
be. If the particles are bound (less energetic) then they are either trapped and oscillate radially in a potential well between two turning
radii or are forced to fall back into the hole by a smaller turning radius that gets infinitesimally closer to the event horizon with the
increase of the angular momentum. \\
In a series of works \cite{abr}, Abramowicz {\it et al.} proposed a general-relativistic concept and definition of a centrifugal force which, in
Schwarzschild geometry, becomes attractive towards the axis of rotation below the radius $r = 3M$ of the first circular photon orbit. This is
not only a general-relativistic effect, as it has a Newtonian analogue with identical geometrical reason \cite{abr}. Extremely compact objects
$(R < 3M)$ contain null geodesics that are captured by the object. De Felice \cite{df} has shown that the effect, which led to the idea of an
attractive centrifugal force, may be explained in a way which preserves the repulsive character of the centrifugal force --- it is argued that
the phenomenon according to which an increase of the angular velocity of an orbiting particle at $r < 3M$ causes more attraction than
centrifugal repulsion has to be interpreted in terms of being below a threshold of a critical gravitational strength, which has its
manifestation in the trapping of the null geodesics, rather than in reversal of the centrifugal force \cite{df}. The situation of trapped
timelike geodesics is similar --- modes with higher angular momentum lack critical gravitational strength to escape the hole. \\
For a Reissner--Nordstr\"om black hole (the centre of which lies in a forbidden region and is thus not accessible by the particle), with the
increase of the angular momentum of an unbound probe, the forbidden regions grow towards the Cauchy horizon and the event horizon respectively
in such way that the radial motion of a probe would only be allowed above the outermost turning radius and, similarly to the Schwarzschild
case, the black hole would be completely shielded for such particles by scattering their geodesics. Similarly, the Hawking radiation would
again be dominated by low angular momentum modes --- the higher the angular momentum of the particle, the closer to the horizons the screen
would be. Additionally in the Reissner--Nordstr\"om case however, if the probe is oppositely charged to the centre, then, depending on the
charges of the probe and the centre, particles with any angular momentum (including zero) and energy within certain range, would be
gravitationally trapped by the forbidden region on the outside of the event horizon. Thus the spectrum of the Hawking radiation would be
significantly voided in this energy region --- that is, but for the quantum tunneling, there would be no particles (charged oppositely to the
centre) coming out of the black hole. The situation of infalling geodesics is analogous --- even the low (and zero) angular momentum modes can
be scattered if the energies are within a certain range. \\
In this article the above effects for Schwarzschild and Reissner--Nordstr\"om black holes are studied in detail and the turning radii are
determined as functions of the probe's conserved energy and angular momentum. Quantum tunneling across the classically forbidden regions is
also analyzed in the framework of the Wentzel--Kramers--Brillouin (WKB) approximation method and the Gamow factor is calculated.

\section{Schwarzschild Black Holes}

The geodesic motion in Schwarzschild geometry (and the considered in next section charged particle motion in Reissner--Nordstr\"om geometry)
are very well studied. Particle motion analysis will be summarized here, following Chandrasekhar's classics \cite{ch}, and further expanded
for finding the forbidden regions. \\
The space-time interval in Schwarzschild geometry (in units $c = 1 = G$) is given by:
\b
ds^2 = - (1 - \frac{2M}{r}) dt^2 + \frac{1}{1 - \frac{2M}{r}} dr^2 + r^2 d\theta^2 + r^2 \sin ^2 \theta \, d\phi^2,
\e
where $M$ is the mass of the centre. \\
As every geodesic can be brought to initially, and hence everywhere, equatorial geodesic by a rotational isometry, attention will be restricted,
without loss of generality, to equatorial geodesics only: $\theta = \pi/2 =$ const. The geodesic equations for a particle of rest mass $m$ are
\cite{carter, fn}:
\b
r^2 \, \frac{dt}{d \lambda} &  =  & \frac{r^2}{r^2 - 2Mr} E r^2, \\
\label{r}
r^2 \, \frac{dr}{d \lambda} & = & \pm \sqrt{E^2 r^4 - (r^2 - 2Mr)(m^2 r^2 + J^2)},  \\
r^2 \, \frac{d\theta}{d \lambda} & =  & 0, \\
r^2 \, \frac{d\phi}{d \lambda} &  =  & J.
\e
where: $\lambda = \tau/m$ is the proper time $\tau$ per unit mass $m, \,\, E$ and $J$ are the conserved energy and the conserved angular
momentum of the particle, respectively. \\
Equation (\ref{r}) governs the radial motion, which can be considered as one-dimensional effective motion:
\b
\frac{1}{2} \dot{r}^2 + V_{\mbox{\tiny eff}}(r) = \frac{\epsilon^2 -1}{2},
\e
where $\epsilon = E/m$ is the specific energy of the three-dimensional motion, $A = (\epsilon^2 -1)/2$ is the specific energy of the
one-dimensional effective motion and the effective potential is:
\b
V_{\mbox{\tiny eff}}(r) = - \frac{M}{r} + \frac{j^2}{2r^2} - \frac{Mj^2}{r^3},
\e
where $j = J/m$ is the specific angular momentum of the particle. \\
Motion is possible only if the specific kinetic energy $\frac{1}{2}\dot{r}^2$ is non-negative.
Particle trajectories are characterized by $\epsilon$ (and, also, by $j$). If $\epsilon < 1$, the particle cannot escape the black hole and the
trajectory is bound. The case of $\epsilon = 1$ corresponds to a particle falling towards the black hole from state of rest at infinity or to a
marginally bound trajectory for an outgoing particle. Finally, the case of $\epsilon > 1$ corresponds to unbound particles. All three cases
will be studied.

\subsection{Marginally Bound Trajectories}

Starting with a marginally bound particle ($\epsilon = 1$), non-negativity of the kinetic energy in the region $r > 0$ necessitates that
$V_{\mbox{\tiny eff}}(r) < 0$, that is:
\b
2Mr^2 - j^2 r + 2 M j^2 > 0.
\e
Thus the kinetic energy is negative between the two turning radii $r_1 \le r_2$ given by:
\b
\label{a_nula}
r_{1,2} = \frac{j^2}{4M} \biggl(1 \pm \sqrt{1 - \frac{16M^2}{j^2}} \, \biggr).
\e
Clearly, for particles with small angular momenta, that is, $j^2 < 16 M^2$, the two turning radii $r_{1,2}$ are not real and therefore the
kinetic energy is always positive and the radial motion --- unimpeded. For larger angular momenta (particles having greater tangential
velocities), a classically forbidden region $(r_1, r_2)$ opens up --- no particle can exist between the two turning radii. Quantum tunneling
is however possible and it will be studied later. \\
The forbidden region is always outside the event horizon $r_H = 2M$. To see this, assume that $r_H > r_1$ and seek a contradiction. The roots
$r_{1,2}$ are real, thus $j^2 \ge 16M^2$. This can be written as $j^2 = (1/\alpha) 16M^2$, where $0 < \alpha \le 1$. The assumption
$r_H > r_1$ leads to the contradiction $\alpha < 0$.

\subsection{Unbound Trajectories}

The specific energy of the three-dimensional motion of unbound particles is $\epsilon > 1$, and the specific energy of the one-dimensional
effective motion of such particles is $A = (\epsilon^2 -1)/2 > 0$. Non-negativity of the kinetic energy leads to:
\b
R(r) \equiv A r^3 + M r^2 - \frac{j^2}{2} r + M j^2 > 0.
\e
At the horizon $r_H = 2M$, the function $R(2M)$ is positive: $R(2M) = 8 A M^3 + 4 M^3 > 0$, since $A > 0$. Thus the horizon is not in a
forbidden region. \\
The cubic equation $A r^3 + M r^2 - \frac{j^2}{2} r + M j^2 = 0$ (in which, again, $A > 0$) always has three roots: $r_1, r_2,$ and $r_3$. As
there are two sign changes in this equation, according to Descartes' rule of signs, there are either two positive roots and one negative root,
or one negative root and two complex roots, or three negative roots. In fact, situation with three negative roots cannot arise as, according to
Vieta's formula, $r_1 r_2 + r_2 r_3 + r_3 r_1 = - j^2/(2A) < 0,$ while each of the products $r_i r_j$ is positive and they cannot add up to a
negative number. As motion occurs for $r >0$, the case of one negative and two complex roots leads to no forbidden region and unimpeded radial
motion. If there are three real roots, one of which, called $r_1$,  is negative and two, called $r_{2, 3}$ respectively (with $r_2 \le r_3$),
are positive, then classically forbidden region exists between the two turning radii $r_2$ and $r_3$. The discriminant of the cubic equation is:
\b
\label{discr}
D = \frac{j^2}{4} [ 2 A (j^2)^2 + (-108 A^2 M^2 - 36 A M^2 + M^2) j^2 - 16 M^4 ].
\e
If $D \ge 0$, there are one negative and two positive roots, if $D < 0$, there are one negative and two complex roots. \\
Forbidden region exists in the first case ($D \ge 0$), that is, when $j^2$ is not between $(j_1)^2$ and $(j_2)^2$, where:
\b
\label{j}
(j_{1, 2})^2 = \frac{(108 A^2 + 36 A - 1)M^2}{4A} \Biggl[ 1 \pm \sqrt{1 + \frac{128 A}{(108 A^2 + 36 A - 1)^2}} \Biggr].
\e
The expression under the root is always positive, thus real $(j_1)^2 \ne (j_2)^2$ always exist. The roots $(j_1)^2$ and $(j_2)^2$ must also be
positive (as they are squares of the angular momentum). The smaller root $(j_1)^2$ is always negative, unless $108 A^2 M^2 + 36 A M^2 - M^2$ is
negative. This is satisfied by $-\frac{1}{6} - \frac{1}{9}\sqrt{3} < A < -\frac{1}{6} + \frac{1}{9}\sqrt{3}$. But $A$ itself is positive for
unbound particles. Therefore, for small values of the specific energy $A$ of the one-dimensional effective motion, i.e. for
$0 < A < -\frac{1}{6} + \frac{1}{9}\sqrt{3}$, or for small values of the specific energy of the three-dimensional motion, i.e.
$\epsilon < \sqrt{\frac{2}{3} + \frac{2}{9}\sqrt{3}}$, the smaller root $(j_1)^2$ is positive. \\
This analysis can be summarized as follows: \\
For small values of the specific energy $A$ of the one-dimensional effective motion, i.e. for $0 < A < -\frac{1}{6} + \frac{1}{9}\sqrt{3}$
and for values of the square of the angular momentum smaller than or equal to $(j_1)^2$ or greater than or equal to $(j_2)^2$, there is a
forbidden region between the two turning radii $r_{2, 3}$ (with $r_2 \le r_3$). If the  square of the angular momentum is between $(j_1)^2$ and
$(j_2)^2$, then forbidden region does not exist and such particles will escape unopposed to infinity or will cross the horizon and hit the
singularity at $r = 0$ if coming from infinity. \\
For values of the specific energy $A$ of the one-dimensional effective motion such that $A \ge -\frac{1}{6} + \frac{1}{9}\sqrt{3}$ [in which
case $(j_1)^2$ is negative and unphysical] and for values of the square of the angular momentum greater than or equal to $(j_2)^2$, there is a
forbidden region between the two turning radii $r_{2, 3}$ (with $r_2 \le r_3$). Particles with square of the angular momentum between 0
[instead of the negative $(j_1)^2$] and $(j_2)^2$ will be unopposed. \\
As seen already, the horizon $r_H = 2M$ is not in the forbidden region [as $R(r) \equiv A r^3 + M r^2 - \frac{j^2}{2} r + M j^2 > 0$ there].
This means that $r_H < r_2$ or $r_H  > r_3$. To see that only the first of these holds, consider the sum of the last two terms in $R(r)$,
namely $- \frac{j^2}{2} r + M j^2$. To the right of the horizon (i.e. for $r > 2M$) and arbitrarily close to it, this is negative for all
$j^2$. On the other hand, the sum of the first two terms in $R(r)$, namely $A r^3 + M r^2 $, is always positive. Therefore, for any specific
energy $A$ of the one-dimensional effective motion, there exists a value of $j^2$ such that $R(r)$ becomes negative just to the right of the
horizon $r_H = 2M$. This means that the inner turning radius $r_2$ is just to the right of the horizon $r_H = 2M$ and $r_2$ approaches $r_H$
for any $A$ as $j^2$ grows. \\
For any energy $A$ and any angular momentum $j$, the root $r_2$ is never greater than $6M$. This can be seen as follows. The specific kinetic
energy $\frac{\dot{r}^2}{2} =  A - V_{\mbox{\tiny eff}}(r) =  A + \frac{M}{r} - \frac{j^2}{2r^2} + \frac{Mj^2}{r^3}$ has a minimum at point
$\frac{j^2}{2M} (1 - \sqrt{1 - \frac{12M^2}{j^2}})$ and $\dot{r}^2$ is negative there in the case of three real roots. Thus $j^2$ must be
greater than $12 M^2$, otherwise there will be no minimum. In the case of three real roots, there is definitely a minimum. In the limit
$j^2 \to 12 M^2$, the local minimum is at a point that tends to $6M$; in the limit of very large $j^2$, the local minimum is at a point that
tends to $3M$. The root $r_2$ is between the horizon $2M$ and the minimum. Thus, the root $r_2$ is never smaller than $2M$ and never greater
than $6M$. \\
To see what values of $j^2$ will guarantee that the root $r_2$ will be arbitrarily close to the horizon $2M$, expand the kinetic energy in
series over the powers of $(r - 2M)$. If $r$ is to be the root $r_2$ (thus $\dot{r}$ will vanish at $r = r_2$), that is very close to $2M$,
truncating the series after the linear term yields:
\b
\label{taylor}
r_2 = 2M \biggl(1 + \frac{\epsilon^2}{1+\frac{j^2}{4M^2}} \biggr).
\e
For the root $r_2$ to be arbitrarily close to $2M$, it is necessary to have $\epsilon^2/[1+ j^2/(4M^2)] \ll 1$ or values of $j$ such that
$j^2 \gg 4M^2 (\epsilon^2 - 1)$. \\
Using this form of $r_2$ and Vieta's formul\ae: $r_1 + r_2 + r_3 = - M/A$, $r_1 r_2 + r_2 r_3 + r_3 r_1 = - j^2/(2A)$, and $r_1 r_2 r_3 = - M j^2/A$,
it is easy to find the outer turning radius $r_3$ for the same range $j^2 \gg 4M^2 (\epsilon^2 - 1)$:
\b
\label{a_plus}
r_3 \simeq \frac{M \epsilon^2}{\epsilon^2 -1}
\Biggl[ -1 + \sqrt{1 + \frac{j^2(\epsilon^2 -1)}{\epsilon^4 M^2 \biggl(1 + \frac{\epsilon^2}{1+\frac{j^2}{4M^2}}\biggr)}} \, \Biggr].
\e

\subsection{Bound Trajectories}

Considering next bound particles (with negative specific energy $A$ of their one-dimensional effective motion,  $A < 0$), the turning points
are the roots of the cubic equation $R(r) \equiv A r^3 + M r^2 - \frac{j^2}{2} r + M j^2 = 0$ which exhibits three sign changes. There are,
therefore, two situations, depending on $j$ for any given $A$. \\
The first one is a case of three positive roots. There are four sub-cases here:  {\it (i)} either one treble root to the right of which
$R(r) < 0$ and motion being impossible, or {\it (ii)} one double root $r_1 = r_2$ and a single root $r_3$ such that $r_1 = r_2 < r_3$ with
$R(r)$ negative to the right of $r_3$ and impossible motion for $r > r_3$, or  {\it (iii)} one single root $r_1$ and a double root
$r_2 = r_3$  greater than $r_1$ with $R(r) \le 0$ to the right of $r_1$ and impossible motion for $r > r_1$, or {\it (iv)} three different
roots $r_1 < r_2 < r_3$ with $R(r)$ negative for $r$ between $r_1$ and $r_2$ and for $r > r_3$ and impossible motion in these intervals. \\
The second situation is a case of two complex roots and a positive root $r_3$. In this case, $R(r)$ is negative for $r$ greater than $r_3$ with
motion being impossible in this region. Note that Vieta's formul\ae $\,$ do not allow a situation with two negative roots and one positive root
to be realized. \\
At the horizon $r_H = 2M$, the sign of the function $R(r)$ depends on $A$. That is, $R(2M) = 4 M^3 (2A + 1)$ is positive for
$-\frac{1}{2} < A < 0$ and negative for $A < - \frac{1}{2}$. \\
Considering the case $A < - \frac{1}{2}$ first, it can be noted that $Ar^3 + Mr^2$ is negative for $r > 2M$. On the other hand,
$-\frac{j^2}{2}r + Mj^2$ is negative for all $j^2$ when $r > 2M$. Thus $R(r)$ never becomes positive for all values of $r > 2M$. Therefore, for
$A < - \frac{1}{2}$, the horizon $r_H = 2M$ is in the forbidden region that extends from $r_3$ to $\infty$. That is, no particles with energy
$A < - \frac{1}{2}$ can exist on the outside of the event horizon. \\
When $- \frac{1}{2} < A < 0$, the horizon is not in a forbidden region [as $R(2M)$ is positive in this case]. Similarly to the situation of
unbound particles, the term $Ar^3 + Mr^2$ is positive to the right of $r_H = 2M$. On the other hand, the other term in $R(r)$, namely
$-\frac{j^2}{2}r + M j^2$, is negative to the right of the horizon for  any $j$. Therefore, for large values of $j^2$, the horizon is just to
the left of a positive root. In case of one positive root $r_3$ only (and the other two --- complex), any particles with energy
$- \frac{1}{2} < A < 0$ will reflect from the forbidden region that extends from $r_3$ to $\infty$ and will fall back into the black hole.
In case of three positive roots with one of them double ($r_1 = r_2 < r_3$ or $r_1 < r_2 = r_3$)  or in the case of a treble positive root
$r_3$, the forbidden region extends from $r_3$ infinity and, again, all particles with energy $- \frac{1}{2} < A < 0$ will fall back into the
black hole. \\
The only remaining situation left to be considered is when $- \frac{1}{2} < A < 0$ and there are three distinct positive roots
$r_1 < r_2 < r_3$. In fact, such situation does not arise for all values of $A$ from $-\frac{1}{2}$ to $0$. For three distinct positive roots
to exist, the discriminant (\ref{discr}) of the cubic polynomial $A r^3 + M r^2 - \frac{j^2}{2} r + M j^2$ must be positive. Given that $A$ is
now negative (but greater than $-\frac{1}{2}$), real positive roots (\ref{j}) for $j^2$ exist if the discriminant of the discriminant $D$
(viewed as a  polynomial in $j^2$) is positive. That is, if $(A + \frac{1}{2}) (A + \frac{1}{18})^3 > 0$. Thus, for
$-\frac{1}{2} < A < -\frac{1}{18}$ there is only one positive root to the right of the horizon $2M$ (and two complex roots) and for
$-\frac{1}{18} < A < 0$ and values of $j^2$ between those determined in (\ref{j}),  there are three positive roots $r_1 < r_2 < r_3$ and two
forbidden regions: between $r_1$ and $r_2$ and to the right of $r_3$. When there are three distinct positive roots, the horizon $2M$ is not in
a forbidden region and it is to the left of $r_1$ for the following reason. The term $-\frac{j^2}{2}r + M j^2$ in $R(r)$ is positive at the
horizon $2M$ or to the left of it for  any $j$. The other term in $R(r)$, namely $r^2 (A r + M)$, is also positive for all values of $r$
between $0$ and the horizon $2M$ and all values of $A$ between $-\frac{1}{18}$ and $0$. Thus, to the left of the horizon, $R(r)$ is everywhere
positive. This means that the horizon can only be to the left of $r_1$ and not between $r_2$ and $r_3$ as in the latter case there would be a
region to the left of the horizon where $R(r)$ is negative. In view of this, particles emitted by the black hole can tunnel across the
forbidden region between $r_1$ and $r_2$. There would be gravitationally trapped particles between $r_2$ and $r_3$ (beyond which there is a
second forbidden region extending all the way to infinity). Black holes cannot evaporate with particles with energy $- \frac{1}{18} < A < 0$.

\subsection{Quantum Tunneling near Schwarzschild Black Holes}

Using the Wentzel-Kramers-Brillouin (WKB) approximation method (see \cite{griffiths}, for example), the transmission coefficient for tunneling
through the classically forbidden region between two turning radii $R_\pm$, in effective potential $V_{\mbox{\tiny eff}}(r)$, with specific
energy $(\epsilon^2 -1)/2$ of the one-dimensional effective motion, will be determined. \\
The one-dimensional Schr\"odinger equation for a particle moving with energy $E$ in potential $U(r)$ can be re-written as:
\b
\frac{d^2 \psi}{dr^2} = - \frac{p^2}{\hbar^2} \psi(r) \, ,
\e
where $p^2(r) = 2m[E - U(r)]$ is the square of the classical momentum of the particle [with $E > U(r)$, so that $p(r)$ is real]. For tunneling
through a potential barrier (namely, across a classically forbidden region between two turning radii $R_\pm$), the WKB-approximated wave
function is given by:
\b
\psi(r) \simeq \frac{C}{\sqrt{\vert p(r) \vert}}\,\, e^{\pm \frac{i}{\hbar}\int\limits_{R_-}^{R_+} \vert p(r) \vert dr} \, ,
\e
where $C =$ const and $\vert p(r) \vert = \sqrt{2m[U(r) - E]}$. \\
The amplitude of the transmitted wave, relative to the amplitude of the incident wave, is diminished by the factor $e^{2 \gamma}$, where
\b
\gamma = \frac{1}{\hbar} \,\, \int\limits_{R_-}^{R_+} \vert p(r) \vert dr.
\e
The tunneling probability $P$ is proportional to the Gamow factor $e^{-2 \gamma}$ \cite{griffiths}. \\
In the present analysis, the forbidden region (corresponding to negative specific kinetic energy $\frac{1}{2}\dot{r}^2$) is where
$V_{\mbox{\tiny eff}}(r) - \frac{\epsilon^2 -1}{2} =  - \frac{M}{r} + \frac{j^2}{2r^2} - \frac{Mj^2}{r^3} - A$ is positive. Thus:
\b
\label{g}
\gamma = \frac{\sqrt{2m}}{\hbar} \,\, \int\limits_{R_-}^{R_+} \,
\sqrt{- Mj^2 \frac{1}{r^3}  + \frac{j^2}{2} \frac{1}{r^2}  - M \frac{1}{r}  - A} \, dr.
\e
As the case of interest is for tunneling across the forbidden region only, the cubic function
$V_{\mbox{\tiny eff}}(r) - \frac{\epsilon^2 -1}{2}$ under the square root in this equation can be approximated by a parabola $W(r)$ that:
{\it (i)} has as roots the roots $R_\pm$ of the cubic function; {\it (ii)} whose maximum occurs where the local maximum of the cubic function
is (this happens  between the roots $R_\pm$, at coordinate $\rho$ which will be determined); and {\it (iii)} whose maxima is equal to that of
the cubic function. That is, near the forbidden region: $V_{\mbox{\tiny eff}}(r) - \frac{\epsilon^2 -1}{2}
= - Mj^2 \frac{1}{r^3}  + \frac{j^2}{2} \frac{1}{r^2}  - M \frac{1}{r} - A  \simeq W(r)
= K \Bigl( \frac{1}{r} - \frac{1}{R_-} \Bigr) \Bigl( \frac{1}{r} - \frac{1}{R_+} \Bigr) =  \frac{K}{r^2 R_+ R_-}(r - R_-)(r - R_+)$, where $K$
is some negative number that will also be determined. \\
As seen earlier, the local maximum of  $V_{\mbox{\tiny eff}}(r) - \frac{\epsilon^2 -1}{2}$ occurs at point:
\b
\rho = \frac{j^2}{2M} \biggl(1 - \sqrt{1 - \frac{12M^2}{j^2}} \, \biggr).
\e
To make sure that the maximum of the approximating parabola is equal to the local maximum of the cubic function and that this happens at the
same point $\rho$, the negative constant $K$ can be easily determined as:
\b
K = \rho^2 R_+ R_- \, \frac{V_{\mbox{\tiny eff}}(\rho) - \frac{\epsilon^2 -1}{2}}{(\rho - R_-)(\rho - R_+)}.
\e
Thus, using the approximating parabola $W(r)$, the factor $\gamma$ given by equation (\ref{g}), becomes:
\b
\gamma & \simeq & \frac{\sqrt{2m}}{\hbar} \, \biggl[ \rho^2 R_+ R_- \,
\frac{V_{\mbox{\tiny eff}}(\rho) - \frac{\epsilon^2 -1}{2}}{(\rho - R_-)(R_+ - \rho)} \biggr]^{\frac{1}{2}}
\,\,\,\, \int\limits_{R_-}^{R_+} \sqrt{\frac{(r - R_-)(R_+ - r)}{r^2}} \,\, dr. \nonumber \\
& = &  \biggl[ \frac{m \pi^2}{2 \hbar^2} \, \rho^2 R_+ R_- \,  \frac{V_{\mbox{\tiny eff}}(\rho) -
\frac{\epsilon^2 -1}{2}}{(\rho - R_-)(R_+ - \rho)} \biggr]^{\frac{1}{2}} (\sqrt{R_+} - \sqrt{R_-})^2.
\e
For marginally bound particles [for which $A = (\epsilon^2-1)/2 = 0$], the limits of integration $R_\pm$ are the two turning radii determined in
(\ref{a_nula}). \\
For unbound particles [for which $A = (\epsilon^2-1)/2 > 0$], the lower limit of integration is $R_- = r_2$, as determined in (\ref{taylor}),
while the upper limit is $R_+ = r_3$, as determined in (\ref{a_plus}). \\
Finally, as seen already, bound particles [for which $A = (\epsilon^2-1)/2 < 0$] would be able to move on the outside of the event horizon only
when $-\frac{1}{2} < A < -\frac{1}{18}$ (in which case they will bounce back towards the horizon by the forbidden region extending all the way
to infinity from the only positive root) or when $- \frac{1}{18} < A < 0$ and values of $j^2$ between those determined in (\ref{j}) --- in
which case there are three distinct positive roots and the particles will encounter an inner forbidden region before the forbidden region
extending to infinity. Quantum tunneling is possible across the inner forbidden region only and the limits of integration in this case will be
$R_-  = 2M [1 + \epsilon^2/(1+j^2/4M^2)]$ and
\b
R_+ =  \frac{(2A + 1)M}{2A} \biggl[ -1 -
\sqrt{1 + \frac{2 A j^2}{(2A + 1)^2 M^2 \biggl(1 + \frac{\epsilon^2}{1+\frac{j^2}{4M^2}}\biggr)}} \, \biggr].
\e
This analysis can, of course, be applied to accreting matter falling towards a black hole. In such case, the forbidden region would lead to the
scattering of incoming unbound or marginally bound particles with large angular momenta. Particles with small angular momenta will not
experience a forbidden region and will cross the horizon. Bound particles with $- \frac{1}{18} < A < 0$ and suitable values of $j$ (resulting
in three distinct roots) will be trapped between $r_2$ and $r_3$, unless they tunnel across to $r_1$ and then cross the horizon. The quantum
tunneling is a competing effect that will diminish the scattering of unbound and marginally bound particles and also the number of trapped
bound particles (excluding the effects of quantum tunneling out of the forbidden region of particles emitted by the black hole).

\section{Unbound Particles in Reissner--Nordstr\" om Geometry}

Reissner--Nordstr\" om geometry in Boyer--Lindquist coordinates (in units $c = 1 = G$) is given by:
\b
ds^2  =  - \frac{\Delta}{r^2} dt^2 + \frac{r^2}{\Delta} dr^2 + r^2 d \theta^2 + r^2 \sin^2 \theta d\phi^2.
\e
where $\Delta  =  r^2 - 2 M r + Q^2$. Here $M$ is again the mass of the centre and $Q$ is the charge of the centre. For a black hole to exist,
$Q^2 < M^2$.  \\
The motion of a particle of mass $m$ and charge $\tilde{q}$ in gravitational and electromagnetic fields is governed by the following equations
\cite{fn}:
\b
r^2 \, \frac{dt}{d \lambda} &  =  &  \frac{r^2}{\Delta} (E r^2  - \tilde{q} Q r) \, ,  \\
\label{rad}
r^2 \, \frac{dr}{d \lambda} & = & \pm \sqrt{(Er^2 - \tilde{q} Q r)^2 - \Delta (m^2 r^2 + J^2)} \, ,  \\
r^2 \, \frac{d\theta}{d \lambda} & =  & 0, \\
r^2 \, \frac{d\phi}{d \lambda} &  =  & J,
\e
where, again, attention is restricted, without loss of generality, to equatorial motion only: $\theta = \pi/2 =$ const. \\
The radial motion is described by equation (\ref{rad}) and, again, it can be considered as a one-dimensional effective motion:
\b
\frac{1}{2} \dot{r}^2 + V_{\mbox{\tiny eff}}(r) = \frac{\epsilon^2 -1}{2}.
\e
This time the effective potential is:
\b
V_{\mbox{\tiny eff}}(r) = \frac{qQ \epsilon - M}{r} \, + \, \frac{j^2 \, + \, Q^2 (1 - q^2)}{2r^2} \,
- \, \frac{Mj^2}{r^3} \, + \, \frac{Q^2 j^2}{2 r^4},
\e
where $q= \tilde{q}/m$ is the specific charge of the particle. \\
Clearly, there are possibilities for the specific kinetic energy
\b
\label{r1}
\frac{1}{2} \dot{r}^2 = \frac{1}{2}(\epsilon - \frac{q Q}{r})^2 - \frac{1}{2}\frac{\Delta}{r^2}(1 + \frac{j^2}{r^2})
\e
to become negative, thus classically forbidden regions exist. For their determination, it will be helpful to study the roots of the quartic
equation $\hat{R}(r) \equiv \frac{1}{2}r^4 \dot{r}^2 = 0$:
\b
\label{r2}
\frac{\epsilon^2 -1}{2} r^4 - (qQ \epsilon - M) r^3 - \frac{1}{2} [j^2 + Q^2 (1 - q^2)] r^2 + M j^2 r - \frac{1}{2} Q^2 j^2 = 0.
\e
Unbound particles ($\epsilon > 1$) will be considered only. As seen previously in the Schwarzschild case, bound particles do not add new
insight. \\
The forbidden region occurs where $\hat{R}(r)$ is negative. In the quartic polynomial (\ref{r2}), the cubic and quadratic terms can change
signs. When the signs of the terms are $+, +, +, +, -,$ respectively, there is only one positive root (together with three negative roots or
with one negative and two complex roots). When the signs of the terms are $+, +, -, +, -,$ or $+, -, +, +, -,$ or $+, -, -, +, -,$ then either
the same situation occurs, or there are three positive roots (together with a negative one). As $\hat{R}(0) = - \frac{1}{2} Q^2 j^2 < 0,$ the
singularity in the centre cannot be reached by an incoming particle as it is in a forbidden region (negative kinetic energy). It should be
noted, that particles with $q^2 > 1$ and zero angular momentum can reach the centre \cite{cg}. \\
It is obvious from (\ref{r1}), that when $\Delta$ is negative, i.e. between the Cauchy horizon $r_-$ and the event horizon $r_+$:
\b
r_\pm = M \biggl(1 \pm \sqrt{1 - \frac{Q^2}{M^2}} \biggr),
\e
then $\hat{R}(r)$ is positive. Thus both horizons are not in a forbidden region. Clearly, in the case of one positive root, motion is impossible
between the centre and the root. Both horizons are to the right of the root. In the case of three positive roots
$\hat{r}_1 < \hat{r}_2 < \hat{r}_3$, there are two forbidden regions --- between the centre and $\hat{r}_1$ and between $\hat{r}_2$ and
$\hat{r}_3$. Quantum tunneling can occur between $\hat{r}_2$ and $\hat{r}_3$. (If one of the three positive roots is double or if there is a
treble positive root, the situation is not too different from the one for the case of a single positive root and will not be studied
separately.) As the horizons $r_\pm$ are not in a forbidden region, they are either to the right of $\hat{r}_3$ (thus all roots being below the
Cauchy horizon $r_-$) or between $\hat{r}_1$ and $\hat{r}_2$. Only the latter situation can be realized for the following reason. The roots of
$\hat{R}(r)$ depend on the parameters of the centre: $M$ and $Q$ and those of the probe: $\epsilon$, $j$, $q$. The parabola
$\Delta = r^2 - 2 M r + Q^2$ is independent on  $\epsilon$, $j$, and $q$ and it has a minimum at $r = M$ which is equal to $-M^2 + Q^2$ and it
is negative. Making $M$ approach 0 (while keeping $Q$ such that $Q^2 <  M^2$) moves the parabola to the left and up. If all three positive
roots were to the left of the Cauchy horizon, making $M \to 0$ (and $Q \to 0$) squeezes all three roots to 0. If $\hat{r}_4$ denotes the
negative root, Vieta's formula $\hat{r}_1 + \hat{r}_2 + \hat{r}_3 + \hat{r}_4 =  2 (q Q \epsilon - M)/(\epsilon^2 -1)$ shows that
$\hat{r}_4$ is finite. Vieta's formula $\sum_{j<k} \hat{r}_j \hat{r}_k = - [j^2 + Q^2(1-q^2)]/(\epsilon^2 - 1)$ gives a finite number on the
right-hand side in the limit $Q \to 0$ and a number approaching zero on the left-hand side.  Therefore it is not possible to have all three
positive roots approach zero and so the roots of the parabola are not above the biggest root $\hat{r}_3$, but are between $\hat{r}_1$ and
$\hat{r}_2$. Thus the forbidden region between $\hat{r}_2$ and $\hat{r}_3$ is above the event horizon $r_+$. \\
The first term $(\epsilon - \frac{q Q}{r^2})^2$ in the expression (\ref{r1}) for the kinetic energy in the case of oppositely charged probe
and centre ($qQ<0$) has a positive lower limit. Then for {\it any} $r $ above the event horizon $r_+$ and for {\it any} energy $\epsilon$ of
the probe, taking a large angular momentum:
\b
\label{jj}
j^2 > r^2 \Bigl[ \frac{r^2}{\Delta} \Bigl(\epsilon - \frac{qQ}{r}\Bigr)^2 - 1 \Bigr]
\e
will result in $\hat{R}(r)$ becoming negative at this point. That is, $\hat{R}(r)$ for particles with such angular momenta (and any energy)
will have three positive roots and these particles will be gravitationally trapped below $\hat{r}_2$ (unless they tunnel across to
$\hat{r}_3$). If $r$ is taken just to the right of the event horizon $r_+$, and as $j^2$ grows large (for any $\epsilon$), $\hat{R}(r)$ will
approach zero just to the right of $r_+$. That is, the larger $j^2$ grows, the closer the root $\hat{r}_2$ will be to the event horizon, i.e.
for large $j^2, \, \hat{r}_2 \to r_+$ for any $\epsilon$. Using similar arguments, it is easy to see that in this case, the smallest positive
root $\hat{r}_1$ will approach the Cauchy horizon $r_-$ for any $\epsilon$ if $j^2$ grows large: $\hat{r}_1 \to r_-$. \\
Considering again (\ref{r1}) in the case of like charged centre and probe ($q Q > 0$), it is obvious that for {\it any} $r $ above the event
horizon $r_+$ and for {\it any} angular momentum $j$ of the probe, taking $\epsilon$ close to $q Q / r$ will result in $\hat{R}(r)$ becoming
negative at this $r$. Of course, as $\epsilon > 1$, such possibility exists only if $|q| > \frac{r}{|Q|}$ [if the specific charge of the probe
does not satisfy this, then $(\epsilon - \frac{q Q}{r})^2$ has a positive lower limit and the situation is analogous to the one considered
above for $qQ < 0$]. This means that for $\epsilon$ close to $q Q / r$, there will be another root above the event horizon and, therefore, it
is not possible to have a single positive root when $qQ > 0$ and $|q| > \frac{r}{|Q|}$ for particles with energy $\epsilon$ in the range
$(\epsilon_1, \epsilon_2)$ where:
\b
\label{ee}
\epsilon_{1,2} = \frac{1}{r} \biggl( qQ \pm \sqrt{\Delta + \frac{j^2}{r^2}} \biggr).
\e
That is, $\hat{R}(r)$ for particles with such energy (and any angular momentum) will have three positive roots and these particles will be
gravitationally trapped below $\hat{r}_2$ (unless they tunnel across to $\hat{r}_3$). Similarly, taking $r$ just under the Cauchy horizon
$r_-$, in the limit $\epsilon \to qQ/r_-$, $\hat{R}(r)$ will approach zero, i.e. the root $\hat{r}_1$ will approach the Cauchy horizon $r_-$
for any $j$ (if the charge of the probe satisfies $|q| > \frac{r_+}{|Q|}$, then it will certainly satisfy $|q| > \frac{r_-}{|Q|}$; otherwise,
if $|q| \le \frac{r_+}{|Q|}$, then the situation is as above for $qQ < 0$). As the probe's specific energy $\epsilon$ cannot have two different
limits at the same time (namely, $\epsilon \to qQ/r_-$ and $\epsilon \to qQ/r_+$), then only one of the roots $\hat{r}_1$ or $\hat{r}_2$ will
tend to the corresponding event horizon at a time. To ensure that both $\hat{r}_1 \to r_-$ and $\hat{r} \to r_+$, it is necessary to take large
angular momentum. These two limits will then hold for any specific energy $\epsilon$. \\
To summarize, for any given specific energy $\epsilon$, only particles with small angular momenta [below the one in (\ref{jj})] will not be
trapped gravitationally and will escape unimpeded to infinity. Alternatively, for any given $j$, particles with energies outside the range
$(\epsilon_1, \epsilon_2)$, as determined in (\ref{ee}), will escape unopposed to infinity. In all other cases, the particles will encounter the
forbidden region between $\hat{r}_2$ and $\hat{r}_3$. Attention will be focused on particles for which the two smaller positive roots of
$\hat{R}(r) = 0$ can simultaneously be approximated by the Cauchy horizon and the event horizon: $\hat{r}_1 \to r_-$ and $\hat{r}_2 \to r_+$,
respectively. As seen, this cannot be achieved by adjusting $\epsilon$ only. It is sufficient however to have particles with large angular
momenta. To determine the lower limit of $j^2$, above which this happens, similarly to the Schwarzschild case, the kinetic energy can be
expanded in power series near the Cauchy horizon and near the event horizon. For the two smaller positive roots of the quartic to be
approximated as $\hat{r}_{2,1} = r_{\pm}  \pm \delta^\pm$, where
\b
\delta^\pm = \frac{r_\pm \Bigl(\epsilon - \frac{qQ}{r_\pm}\Bigr)^2}{2 \Bigl(1 + \frac{j^2}{r_\pm} \Bigr)\sqrt{M^2 - Q^2}
\mp 2 q Q \Bigl(\epsilon - \frac{qQ}{r_\pm} \Bigr)}
\e
are sufficiently small, it is necessary that the angular momentum satisfies:
\b
j^2 \gg r_\pm \Biggl[\pm \frac{(\epsilon r_\pm)^2 - (qQ)^2}{2\sqrt{M^2 - Q^2}}  - r_\pm\Biggr],
\e
thus guaranteeing that, for any energy $\epsilon$,  $\hat{r}_1 \to r_-$ and $\hat{r}_2 \to r_+$ will hold. Note that $\delta^\pm$ are both
positive (otherwise an expansion of the kinetic energy near one horizon will be "reaching across" the other horizon for roots). There is always
a root between $r=0$ and $r_-$. If, however, $\delta^+$ is only negative, it means that there will be just one positive root (below the Cauchy
horizon $r_-$) and no forbidden region beyond $r_+$. In the case of oppositely charged probe and centre, $\delta^+$ is positive and thus there
always is a root to the right of $r_+$. If the two charges have the same signs and $\delta^+$ is only negative, then there will be no root to
the right of the event horizon. This gives an inequality involving the parameters of the model. Solving it, for example, for the specific
energy $\epsilon$, then there is just one positive root if:
\b
\epsilon \ge \frac{qQ}{r_+} + \frac{\sqrt{M^2 - Q^2}}{qQ} \Bigl( 1 + \frac{j^2}{r_+^2} \Bigr).
\e
Similarly to the Schwarzschild case, in the limit $\hat{r}_1 \to r_-$ and $\hat{r}_2 \to r_+$, the upper boundary of the forbidden region,
i.e. the biggest positive root $\hat{r}_3$ can easily be found using Vieta's formul\ae \, $\hat{r}_1 + \hat{r}_2 + \hat{r}_3 + \hat{r}_4
=  2 (q Q \epsilon - M)/(\epsilon^2 -1)$ and $\hat{r}_1 \hat{r}_2 \hat{r}_3 \hat{r}_4 = - Q^2 j^2/(\epsilon^2 - 1)$ by substituting the roots
$\hat{r}_1$ and $\hat{r}_2$, as determined above, and eliminating the negative root $\hat{r}_4$. This gives:
\b
\hat{r}_3 & \!\!\! \simeq & \!\!\! \Bigl[ M + \frac{qQ\epsilon - M}{\epsilon^2 - 1} + \frac{\delta^+ - \delta^-}{2}  \Bigr]
 \nonumber \\
& & \hskip-1.22cm \times \! \Biggl[ -1 + \!\! \sqrt{1 \!
+ \frac{Q^2 j^2}{(\epsilon^2 - 1) (Q^2 - \delta^+\delta^- - r_+ \delta^- + r_- \delta^+)
\Bigl[ M + \frac{qQ\epsilon - M}{\epsilon^2 - 1} + \frac{\delta^+ - \delta^-}{2}  \Bigr]^2  }}\Biggr]\!. \nonumber \\
&&
\e
As in the Schwarzschild case, it will be useful to approximate the effective potential between the roots  $\hat{r}_2$ and $\hat{r}_3$ by the
parabola $\hat{W}(r)$. This parabola will have $\hat{r}_2$ and $\hat{r}_3$ as roots. It is however difficult to find the local maximum of
$V_{\mbox{\tiny eff}}(r) - \frac{1}{2}(\epsilon^2 - 1)$ (as it is a root of a cubic equation) and thread the parabola through it. To avoid this
difficulty, one can take the midpoint $\hat{\rho} = (\hat{r}_3 - \hat{r}_2)/2$ between the root $\hat{r}_2$ and the root $\hat{r}_3$ (i.e. the
midpoint of the forbidden region) and make the approximating parabola go through point of coordinates $(\hat{\rho}, \hat{R}(\hat{\rho}))$. The
parabola is therefore:
\b
\hat{W}(r) = \hat{K} \Bigl( \frac{1}{r} - \frac{1}{\hat{R}_-} \Bigr) \Bigl( \frac{1}{r} - \frac{1}{\hat{R}_+} \Bigr)
=  \frac{\hat{K}}{r^2 \hat{R}_+ \hat{R}_-}(r - \hat{R}_-)(r - \hat{R}_+),
\e
where $\hat{R}_- = \hat{r}_2,  \,\, \hat{R}_+ = \hat{r}_3,$ and $\hat{K}$ is the negative constant:
\b
\hat{K} = \hat{\rho}^2 \hat{R}_+ \hat{R}_- \, \frac{V_{\mbox{\tiny eff}}(\hat{\rho}) - \frac{\epsilon^2 -1}{2}}{(\hat{\rho} -
\hat{R}_-)(\hat{\rho} - \hat{R}_+)}.
\e
For unbound particles in Reissner--Nordstr\"om geometry, the factor $\gamma$ given by equation (\ref{g}), is:
\b
\gamma & \simeq & \frac{\sqrt{2m}}{\hbar} \, \biggl[  \hat{\rho}^2 \hat{R}_+ \hat{R}_- \, \frac{V_{\mbox{\tiny eff}}(\hat{\rho}) -
\frac{\epsilon^2 -1}{2}} {(\hat{\rho} - \hat{R}_-)(\hat{R}_+ - \hat{\rho})} \biggr]^{\frac{1}{2}}
\,\,\,\, \int\limits_{\hat{R}_-}^{\hat{R}_+} \sqrt{\frac{(r - \hat{R}_-)(\hat{R}_+ - r)}{r^2}} \,\, dr. \nonumber \\
& = &  \biggl[ \frac{m \pi^2}{4 \hbar^2} \,  (\hat{r}_3 - \hat{r}_2)^2 \hat{r}_2 \hat{r}_3 \, \frac{V_{\mbox{\tiny eff}}(\hat{\rho}) -
\frac{\epsilon^2 -1}{2}} {(\hat{r}_3 - 3 \hat{r}_2)(\hat{r}_3 + \hat{r}_2)} \biggr]^{\frac{1}{2}} (\sqrt{\hat{r}_3} - \sqrt{\hat{r}_2})^2.
\e
It is interesting to consider the case of a neutral particle ($q = 0$) in Reissner--Nordstr\"om geometry. From equation (\ref{r1}) it is clear
that for any value of the particle's energy $\epsilon$, at any point $r$ above the event horizon $r_+$, a large angular momentum (above
$r^4 \epsilon^2 / \Delta$) will lead to gravitational trapping and possibility for quantum tunneling across the forbidden region. Neutral
particles cannot reach the singularity.

\section{Conclusions}

Radial timelike motion near Schwarzschild and Reissner--Nordstr\"om black holes exhibits interesting properties. In the former case, higher
angular momentum of a probe leads to a forbidden region opening up on the outside of the event horizon and the higher the angular momentum, the
closer to the event horizon the forbidden region is. The width of the region grows unboundedly with the increase of the angular momentum and
thus a Schwarzschild black hole will be completely "sealed" above the horizon from either side of the potential barrier for high angular
momentum particles (by their screening and scattering, respectively) --- even the quantum tunneling across would cease working with the
widening of the forbidden region. \\
The picture in the Reissner--Nordstr\"om case is richer: the same effects hold in relation to high angular momentum modes and, additionally,
zero or low angular momentum modes, charged oppositely to the centre, exhibit the same behaviour if their energies are within a certain range.
In the Reissner--Nordstr\"om case too, suitable values of the angular momentum or energy can "seal" the black hole and the widening of the
forbidden region can exclude the quantum tunneling effects. The energy spectrum of the emitted particles (or those that fall into the hole)
would have a characteristic depleted range. For both cases, only low angular momentum modes would travel unimpeded.

\section*{Acknowledgements}

It is a pleasure to acknowledge discussions with Elena G. Tonkova and Milena E. Mihaylova.

\end{document}